\documentclass[11pt,twoside]{article}

\usepackage{asp2006}
\usepackage{fancyhdr,graphicx,caption,rotating,multirow,lscape}
\DeclareGraphicsExtensions{.eps,.eps.gz,.ps,.jpg,.tiff}

\begin{document}

\title{Potential of Photometric Searches for Transiting Planets }


\author{Fr\'ed\'eric Pont}
\affil{Geneva University Observatory, Switzerland}
\author{{\bf The ISSI Working Group on Transiting Planets:} Suzanne Aigrain, Gaspar Bakos, Timothy Brown, Andrew Collier Cameron, Hans Deeg, Anders Erikson, Tristan Guillot, Keith Horne,  Alain Lecavelier, Claire Moutou, Heike Rauer, Didier Queloz, Sara Seager}



\begin{abstract} 
Many ground-based photometric surveys are now under way, and five of them have been successful at detecting transiting exoplanets. Nevertheless, detecting transiting planets has turned out to be much more challenging than initially anticipated. Transit surveys have learnt that an overwhelming number of false positives and confusion scenarios, combined with an intermittent phase coverage and systematic residuals in the photometry, could make ground-based surveys rather inefficient in the detection of transiting planets. We have set up a working group on transiting planets to confront the experience of the different surveys and get a more complete understanding of these issues, in order to improve the observing strategies and  analysis schemes for ongoing surveys, and to prepare for the coming Corot and Kepler  space missions. This contribution presents the current results of our working group. 
\end{abstract}


\section{Enter the transiting planets, one by one}

Transiting planets play a fundamental role in the exploration of planets outside the Solar System. The transit lightcurve,  when combined with a radial velocity orbit, gives direct access to the planet's mean density, a fundamental physical parameter. Spectroscopy during the transit and infrared photometry of the passage of the planet behind the star have already permitted direct measurements of some spectral characteristics of the atmosphere of extrasolar gas giants- an achievement that would have sounded futuristic only a decade ago.

\begin{figure}[!th]
\centering
\begin{minipage}[c]{.49\textwidth}
\includegraphics[angle=-90,width=6.7cm]{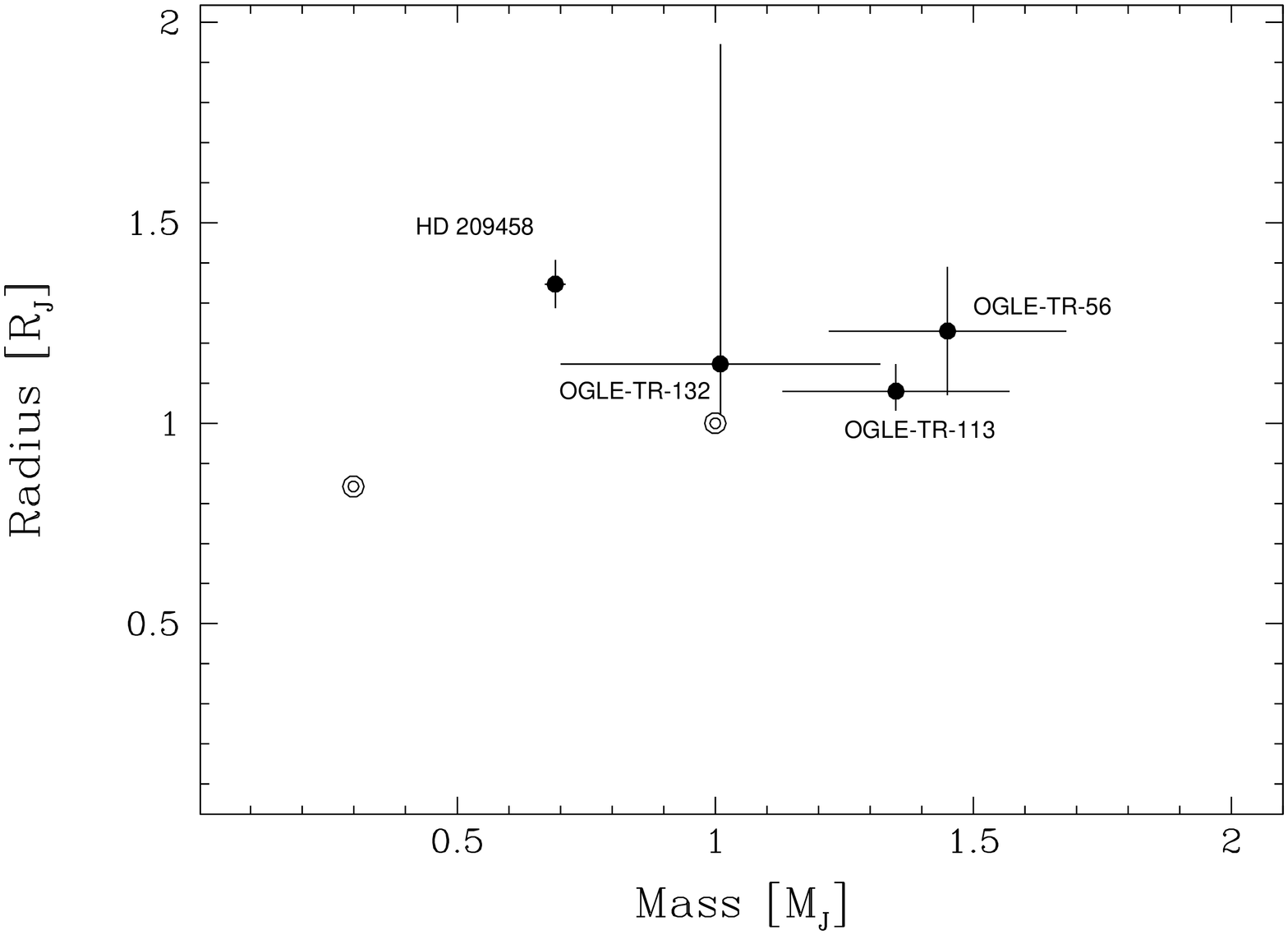}
\end{minipage}
\begin{minipage}[c]{.49\textwidth}
\includegraphics[angle=-90,width=7cm]{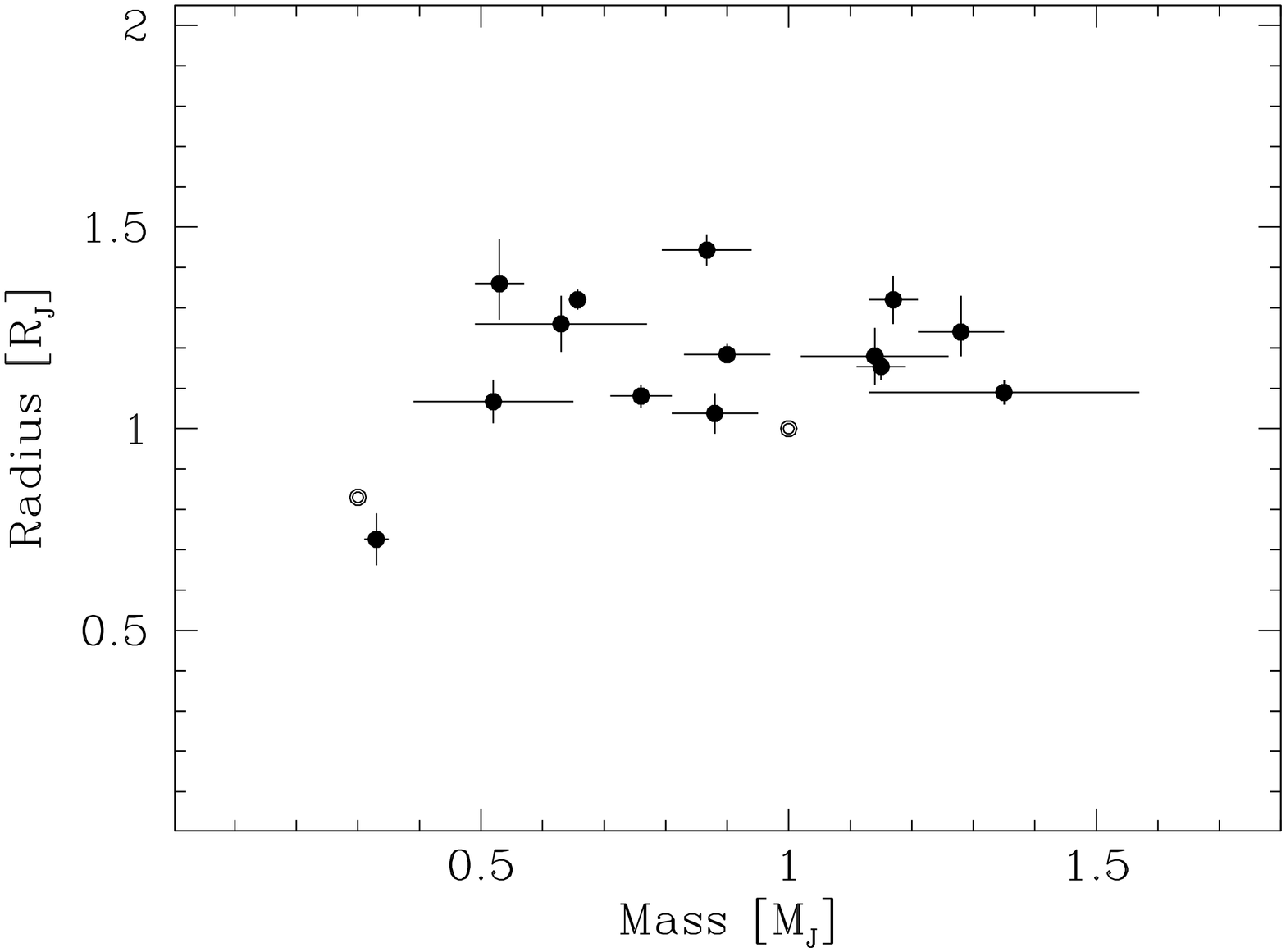}
\end{minipage}
\caption{ {\bf Left:} 
the first empirical mass-radius relation for extrasolar planets, from Bouchy et al. (2004). The symbols indicate Jupiter, Saturn and the four transiting hot Jupiters then known. {\bf Right:} the gas giants mass-radius diagram as of November 2006. Individual values and references can be found on {\small obswww.unige.ch/$\sim$pont/TRANSITS.htm}.}
\label{mrr}
\end{figure}

The quest for transiting exoplanets is a nascent field. The first mass-radius diagram for hot Jupiters was published two years ago (Fig.~\ref{mrr}), with only four hot Jupiters and considerable error bars. As of November 2006, fourteen transiting hot Jupiters are known. Indeed, the Heidelberg meeting proved extremely well-timed, with four detections announced within a month of the meeting, including two in this volume (Street et al.). For many participants, the Heidelberg workshop will have marked the transition from a period of doubts about the potential of ground-based transit surveys, to a heady anticipation on the accelerating rate of discoveries. 


\section{``Hot Jupiters galore''? Not quite}

The discovery of the first close-in extrasolar planet in 1995, then the first transiting hot Jupiter in 1999, prompted several teams to attempt the photometric detection of planetary transits. The procedure seemed temptingly straightforward: Doppler surveys have shown that hot Jupiters orbit about 1\% of solar-type stars, and their transiting probability is near 10\%, so that one star in a thousand should exhibit periodic dimming at the $\sim$1\% level due to transits by a close-in gas giant.  Modern CCD detectors have made the monitoring of several thousand stars at the $\sim$1\% level routine.

In a review entitled ``hot Jupiters galore'' that captured the imagination of many, Horne (2003) predicted with order-of-magnitude simulations that current or planned transit surveys had the capacity of detecting scores of transiting planets per month. In reality, however, only the most ambitious surveys have yielded any detections, and then at a rate at least one order of magnitude lower than predicted. Rather than one in a thousand, the detection rates have hovered near one in several tens of thousands. Simple scaling arguments show that dozens of transiting hot Jupiters present in the data were not detected.

Last year, we have set up a  Working Group on transiting planets in the framework of the International Space Science Institute in Bern (Switzerland), to explore the cause of the mismatch between expectation and actual results for ground-based transit surveys. The ISSI group brings together members of several major transit surveys ({\small OGLE, TrES, SWASP, HAT, COROT, BEST, MONITOR, EXPLORE}) as well as specialists in radial velocity searches and spatial transit measurements. We present here the current results of the team on the specific issue of the potential of planetary transit surveys.

\section{All that glitters is not gold}

Photometric transit surveys identify objects which exhibit short periodic dimming of their luminosity with the appropriate shape, but there can be several causes to this signal other than a transiting planet. Spectroscopic follow-up has turned out to be the most powerful diagnostic to reveal the real nature of a transit candidate, either by showing that the target star has a spectral type incompatible with the planetary interpretation, or by measuring the orbital motion of the system, and therefore the stellar or planetary mass of the body producing the signal. It has been a common experience of all photometric searches for planetary transits that stellar ``impostors'' outnumber real planetary transits by more than one order of magnitude. 

Since 2002, some of us have been studying the transit candidates of the OGLE survey and the results have shed much light on the nature of the difficulties encountered by photometric transit searches. The OGLE survey has been the one publishing the most candidates and producing the most solved cases of planetary transits and stellar mimics (Udalski et al.  2004 and references therein). It uses a 1.3m telescope at Las Campanas Observatory in Chile with a 8k$\times$8k mosaic camera to monitor fields of about one square degree near the Galactic plane during 3-4 months each year. The targets typically have $I$ magnitudes between 14 and 17. The OGLE team has published 177 shallow transit candidates from three observing seasons. A large fraction of these candidates were monitored with the HARPS/3.6m and UVES/VLT spectrographs at ESO (Bouchy et al. 2005, Pont et al. 2005a, +work in progress).  This effort has permitted the identification of five transiting planets, and also an empirical view of the panorama and frequency of the four main different types of  ``impostors'': 
(1) Transit of a solar-type star by an M dwarf companion; 
(2) Eclipsing binary diluted by a third star;
(3) Grazing eclipse of two solar-type stars;
(4) Intrinsic or measurement luminosity fluctuation.

Planets and possible but unconfirmed  planets represent about 5\% of the total. 


This type of panorama of planetary transit mimics is typical in ground-based surveys. There are small differences -- for instance deeper surveys encounter more physical triple systems in the second category while shallow wide-angle surveys find more eclipsing binaries diluted by background giants -- but the overall pattern has proved very similar in all surveys.

The fact that impostors outnumber planetary transits by a large margin has major implications for photometric transit surveys. Follow-up observations become a necessary component of the survey. Transit ``impostors'' have proven surprisingly imaginative at imitating planetary transits to a remarkable degree - short of a clear radial velocity planetary orbit (e.g. Pont et al. 2005b, Mandushev et al. 2005). As a result, data excluding some mimic scenarios but not others are not enough to gain sound confidence in a transiting planet detection. In shallow small-telescope surveys, the candidates can be sorted out with follow-up observations from mid-sized telescopes. In medium-deep surveys like OGLE, the follow-up implies a major investment in large-telescope time. In deeper surveys, the consequences are even more drastic: candidates cannot be confirmed, and since most of them may be impostors, no scientifically useful conclusions can be drawn from the candidates. For this reason, deep surveys such as {\small EXPLORE/OC} and Monitor on open clusters, and the HST survey in the Galactic bulge, have limited potential for extrasolar planet studies.

The presence of impostors can delay the detection of transiting planets in a survey or make it more difficult. It does not, in itself, diminish the potential of a survey (but for the important caveat that the targets have to be bright enough for clear radial velocity confirmation). We now turn to a major possible cause of the detection of fewer planets than predicted.

\section{Pink noise on rosy expectations}

Achieving photon-noise limited photometry on many thousands of stars simultaneously at the percent level and below is no easy task.  A whole host of instrumental and atmospheric effects can disturb the photometric measurements, effects usually lumped under the name of ``systematics''.  The presence of these systematics is an obvious obstacle to transit detection, and a large part of the effort in photometric data reduction is devoted to mitigating them (e.g. with advanced decorrelation algorithms like "TFA", Kovacs et al. 2005; "SysRem", Tamuz et al. 2005). Only recently, however, has the specific impact of these systematics on the potential of transit surveys been studied directly. In Pont, Zucker \& Queloz (2006, hereafter PZQ), we presented the calculations that form the basis of this aspect of the work of the ISSI working group.

We were first alerted to this issue by the presence of some objects among the OGLE candidates that did not seem to present real transit-like signals, and by the strangeness of the five OGLE transiting planets. The presence of ``false positives'' (targets without real transit signal) indicated that the OGLE team had done a thorough job in the census of candidates and gone as near as the detection threshold as possible. The characteristics of the planets, on the other hand, indicated that the efficiency of the survey was low in terms of hot Jupiter detections. The five detected hot Jupiters  all either have anomalously short periods ($P<2$ days, an exceedingly rare case among radial velocity planets), or periods exact multiples of 1 or {\small 1/2} Earth day, or exceptionally deep transits. Most of the candidates actually accumulated two of these factors that strongly favour detection (see Table~\ref{transits}). The transiting planets detected by other photometric surveys also share these peculiar features.

Simple arguments indicate that there should be around 50 to 100 detectable transiting hot Jupiters among the OGLE targets. Therefore the OGLE survey, despite its impressive success, also confirms the ratio of one order of magnitude between projections and real detections. 

An obvious factor affecting the number of detections is the fact that observations can only be gathered during the night. Some transits are simply missed because they occur during the day, or in a night when no observations were taken. This effect is simple to calculate, and for the OGLE survey, which gathered data typically during 60 nights per season, it is expected to prevent less than half of hot Jupiter detections. 

The second limiting factor is more subtle. Because planets produce rather shallow transits, many candidates will be close to the detection threshold, so that the exact position of this threshold will have a strong impact on the number of planets detected. To first order, a planetary transit signal is a box-shaped drop in flux, so that the significance of a detection is generally approximated by the discrepancy of the mean flux during the transit compared to the flux outside the transit, expressed in units of standard deviations. In other words:\\
$S_d =  \sqrt{n}\,  d/\sigma$\\
where $S_d$ is the transit significance statistic, $d$ is the transit depth, $n$ the number of data points during the transit and $\sigma$ the mean photometric error. Mathematical arguments indicate that a threshold of $S_d \simeq 7$ is applicable to photometric transit surveys (e.g. Jenkins et al. 2002). 

\begin{figure}[h]
\centering
\begin{minipage}[c]{.49\textwidth}
\includegraphics[width=7cm]{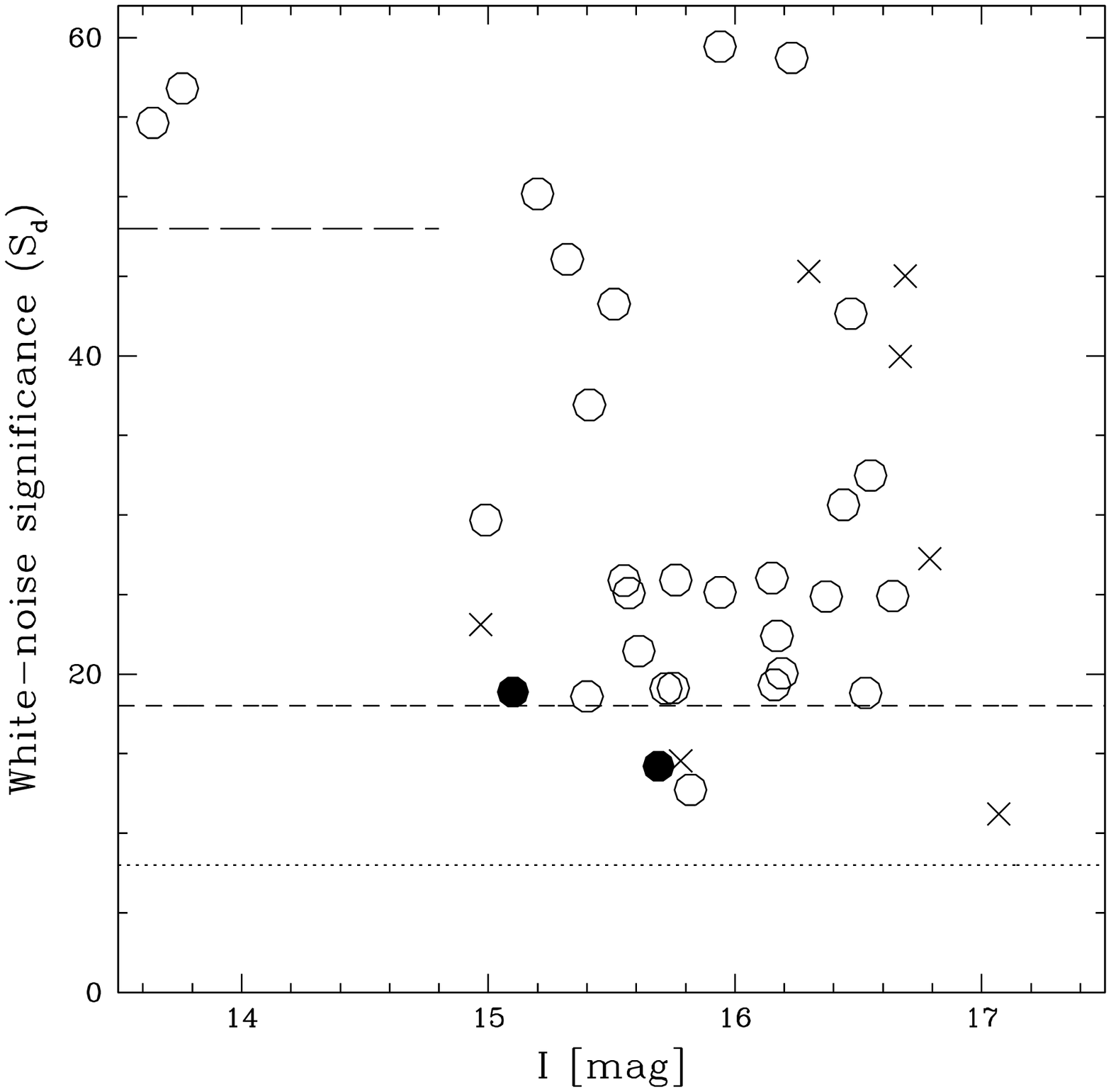}
\end{minipage}
\begin{minipage}[c]{.49\textwidth}
\includegraphics[width=7cm]{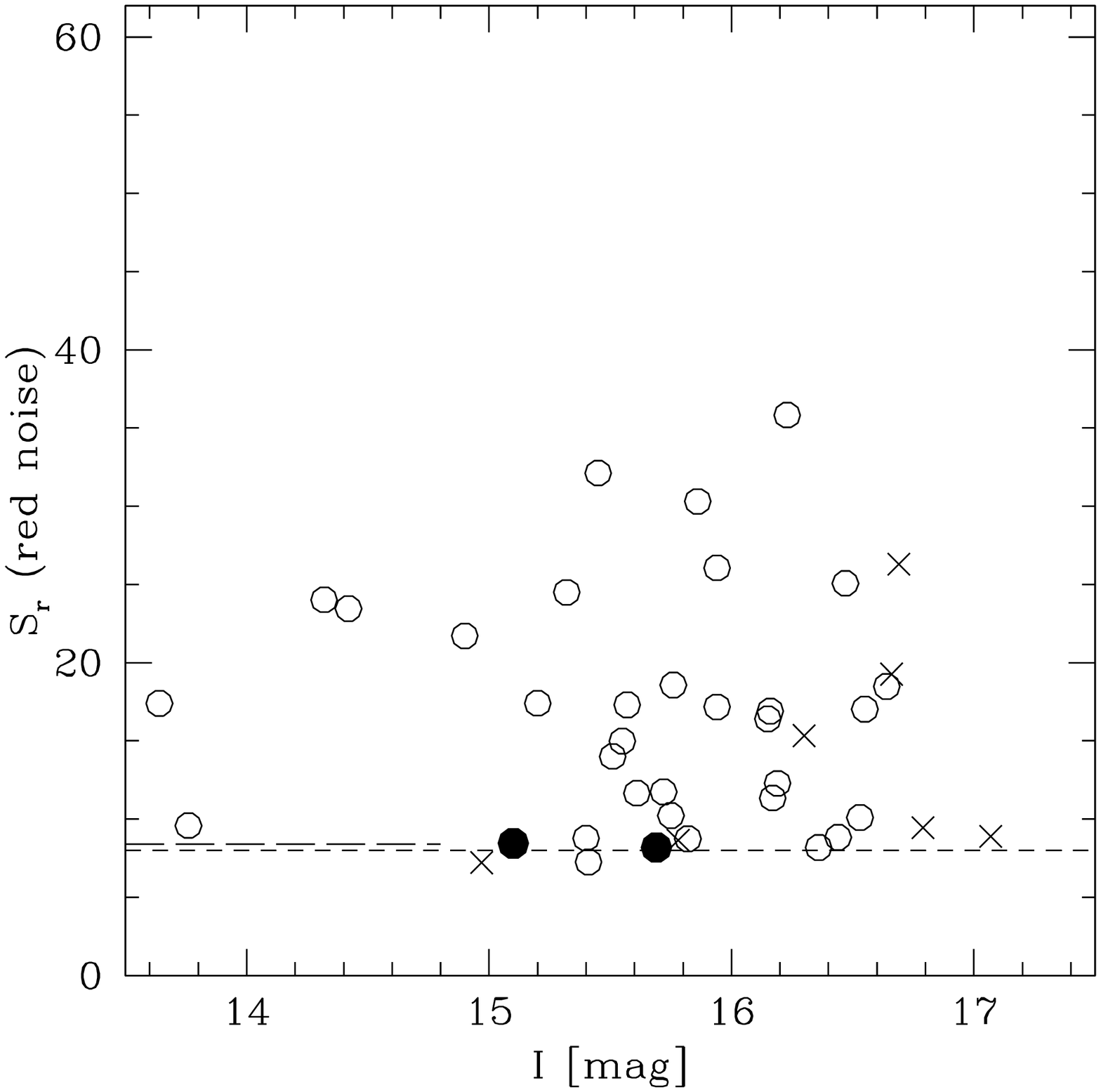}
\end{minipage}
\caption{ Significance statistics versus magnitude for the OGLE transit candidates in Carina. {\bf Left panel:} significance assuming uncorrelated noise. {\bf Right panel:} Significance taking to account the correlation in the noise (red+white, or ``pink'', noise). $\circ$: confirmed eclipsing binaries or transiting planets; $\bullet$  suspected statistical false positives; $\times$: other candidates that could be false positives. The lines indicate several threshold levels mentioned in the text.
\label{snr}}
\end{figure}

The left panel of Figure~\ref{snr} shows the distribution of $S_d$ for the OGLE candidates. The crosses and filled circles indicate targets that may be statistical false positives. What this figure shows is that the actual threshold defined by the OGLE candidates is much higher than $S_d \simeq 7$. In fact, there is only one confirmed signal below $S_d=18$. For brighter magnitudes, there is even no candidate with $S_d<48$.

\begin{figure}[!ht]
\centering
\includegraphics[angle=0,width=9cm]{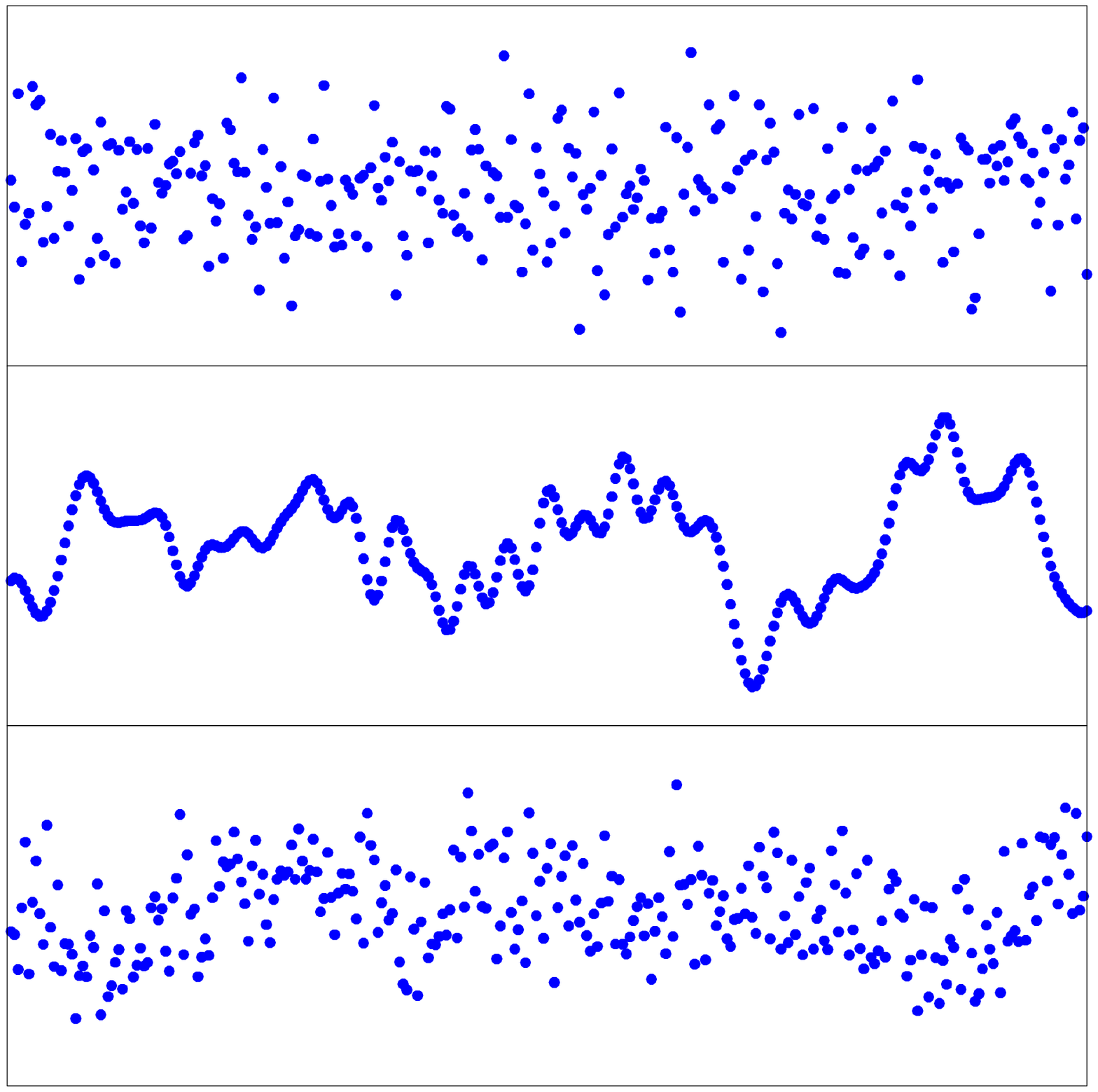}
\includegraphics[angle=0,width=9cm]{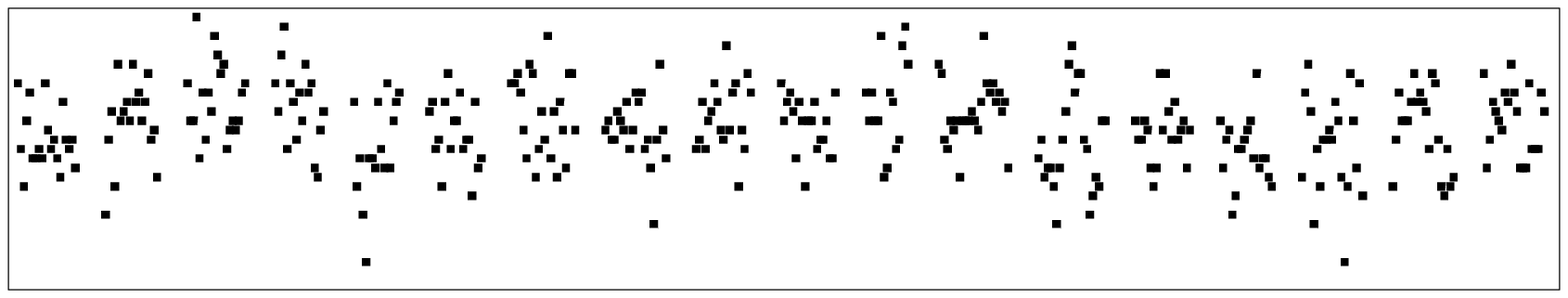}
\caption{{\bf Top:} Example of white, red, and ``pink'' (red+white) noise respectively. With the same variance, transit signals are much harder to detect in the presence of red noise. {\bf Bottom:} a real transit search lightcurve (from the OGLE survey) spanning 18 nights of measurements. The day intervals have been compressed for the display. 
Correlated noise is apparent on several timescales. at the level of a few millimagnitudes. 
 \label{pink}}
\end{figure}

The fundamental reason for this puzzling behaviour is that the expression of $S_d$ above rests on the assumption of uncorrelated noise. However, photometric systematics introduce in the data a noise that is strongly correlated, and especially effective at producing correlated fluctuations -- ``red noise'' in signal analysis jargon -- with the same timescale as planetary transits. Many parameters affecting ground-based photometry, like airmass, seeing, temperature, telescope tracking, vary gradually over timescales of one or a few hours. These effects add a red noise component to the white photon noise, and make photometric noise ``pink'' (See Figure~\ref{pink}). Such noise can produce transit-like features that require the actual detection threshold to be placed much higher than could be the case with purely uncorrelated noise .

In PZQ, we reconsider the detection significance statistic in the presence of correlated noise. We introduce a ``red'' significance indicator, $S_r$, that takes into account the observed correlation in the photometric noise. $S_r$ can be much smaller than $S_d$ in transit surveys, especially for the brighter objects.

 The right panel of Fig.~\ref{snr} shows the distribution of $S_r$ for the same objects as the left panel. This time, a threshold not far above 7 (actually in the 8-9 range) is found to be a good description of the actual detection limit. 

This difference in actual threshold due to correlated noise usually results in a significant or even drastic reduction in the detection potential of transit surveys. For instance, if a given transiting planet  needs 3 measured transits to reach $S_d>7$,  it will require 20 transits covered to reach $S_d>18$ and $S_r>8$ (the effective threshold seen in the OGLE survey).

The work of the ISSI working group has indicated that the correlated noise had an analogous effects in all transit surveys examined (next Section). Values of $S_d$ and $S_r$ are given for transiting planets in Table~\ref{transits}, showing that values of $S_d$ are always far higher than the theoretical threshold of 7, whereas the distribution of $S_r$ makes much better sense.

\begin{table}[!ht]
\smallskip
\caption{Transiting exoplanets detected at the time of writing by photometric surveys involved in the ISSI working group. $S_d$ and $S_r$ are the significance statistics without/with accounting for red noise. Note the short and resonant periods, and the depths larger than 1\%.}
\begin{center}
{\small
\begin{tabular}{lcccc|lcccc}
\tableline
\noalign{\smallskip}
Name & Period & Depth & $S_d$ & $S_r$& Name & Period & Depth & $S_d$ & $S_r$ \\
         & [days] & [\%] & & &      &   [days] & [\%] &  &      \\
\noalign{\smallskip}
\tableline
\noalign{\smallskip}
OGLE-TR-10 & 3.1& 1.2& 25 & 9  &      TrES-1 & 3.0& 1.9& 30 & 14      \\
OGLE-TR-56 & 1.2& 1.1 & 14 & 13 &     TrES-2 & 2.5& 1.6& 37  & 12      \\
OGLE-TR-111 & 4.0& 1.8& 21 & 15  &   HAT-P-1 & 4.5 & 0.6 & 14 & 9     \\ 
OGLE-TR-113  & 1.4& 2.1&38  & 24  &   WASP-1 & 2.5& 1.1& 18 & 11       \\ 
OGLE-TR-132  & 1.7 & 0.8& 19 & 12  &   WASP-2 & 2.2& 1.8 & 21 & 17       \\ 
\noalign{\smallskip}
\tableline
\end{tabular}
}
\end{center}

\label{transits}
\end{table}

\section{The ISSI group surveys surveys}

\begin{figure}[!th]
\centering
\begin{minipage}[c]{.49\textwidth}
\includegraphics[angle=-90,width=7cm]{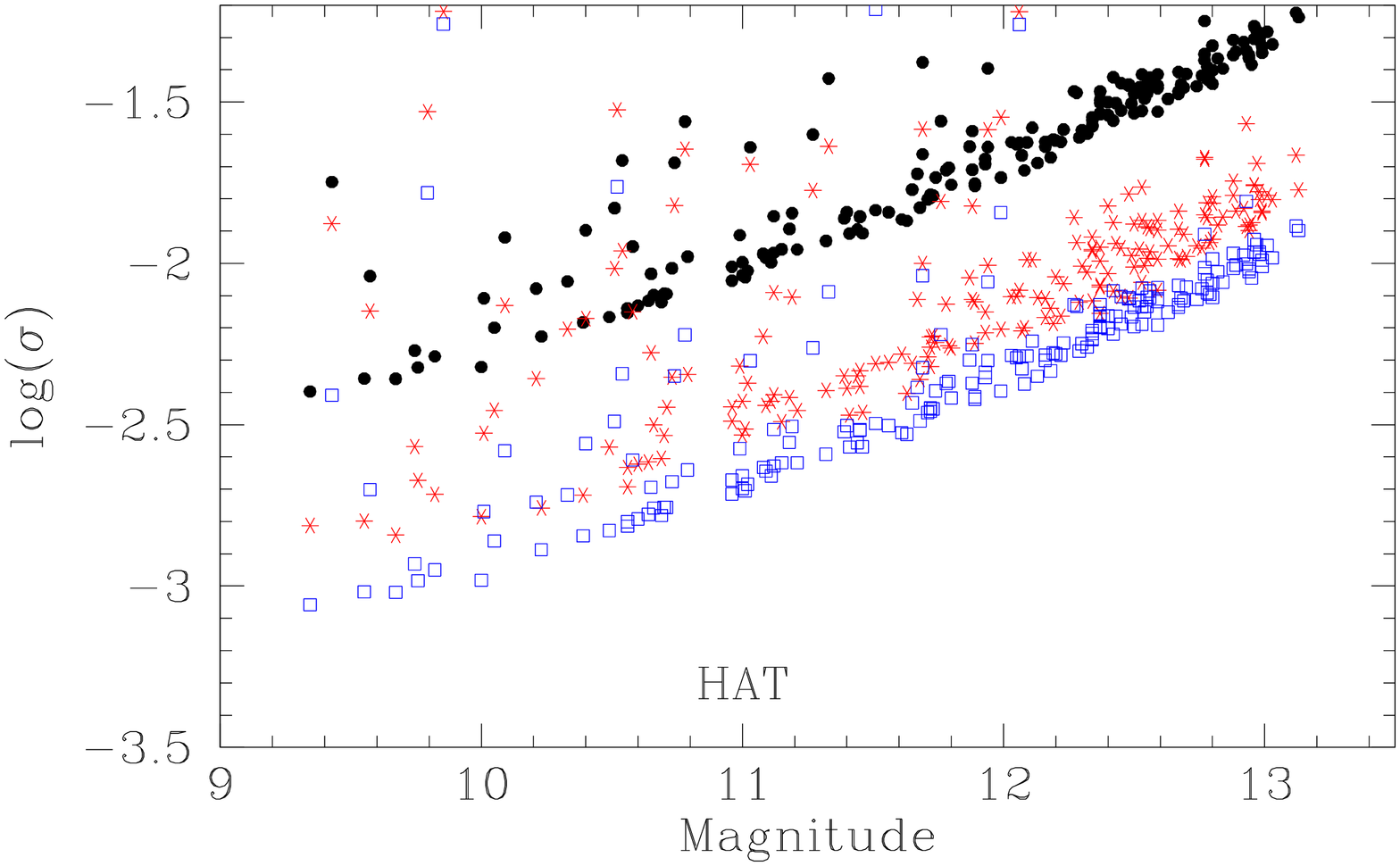}
\end{minipage}
\begin{minipage}[c]{.49\textwidth}
\includegraphics[angle=90,width=6.5cm]{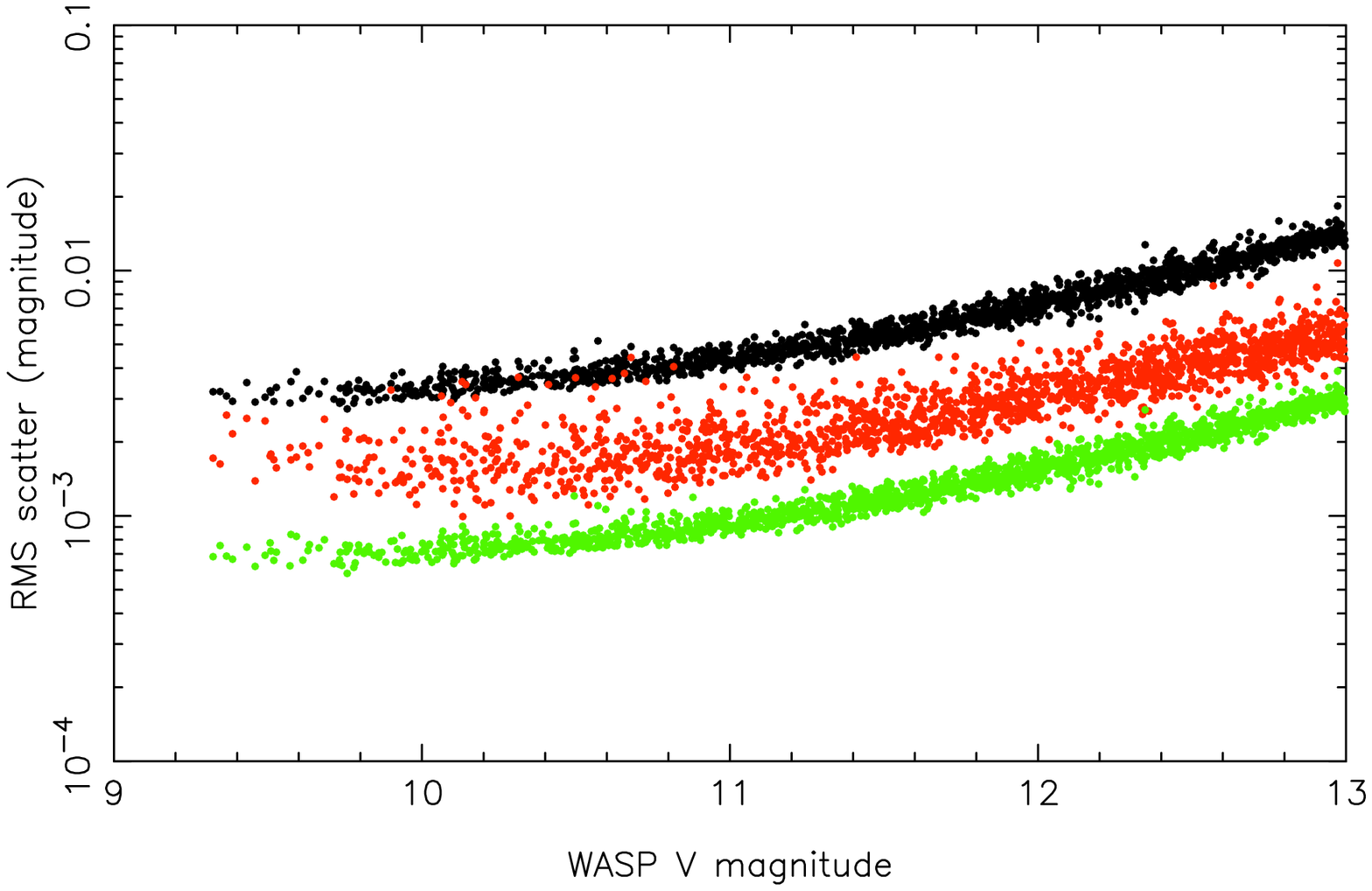}
\end{minipage}
\caption{Two example of the ``red noise plots'' used for the comparison of surveys by the ISSI team, for HAT (left) and SuperWASP (right) data. The upper group of symbols shows the rms of lightcurves for each target, the middle group shows the rms of means over transit-length time intervals, and the lower group the expected rms for uncorrelated noise. SuperWASP plot from Collier Cameron et al. (2006).}
\label{magsig}
\end{figure}

The ISSI working group has taken up the task of comparing the amplitude and behaviour of correlated noise in several different ground-based transit surveys. We have first settled on a standardized diagnostic for the red noise. Two examples are given in Figure~\ref{magsig}. We start from the standard plot used to illustrate the performance of wide-field photometry: the standard deviation of individual lightcurves as a function of the target magnitude. To this we add for each lightcurve the position of the {\it standard deviation of the mean magnitude over a transit-length time interval} (typically 2-3 hours). We also plot the theoretical position of this standard deviation in the case of  uncorrelated noise, which is simply $\sigma / \sqrt{n}$ if the uncertainties are constant (where $n$ is the number of data points in the transit-length interval). For purely uncorrelated noise, the last two sets of points should overlap. The amount by which the second set exceeds the first indicates the amplitude of the red noise over a transit-length timescale.

In PZQ, we introduced a parameter called $\sigma_r$, to characterise the amplitude of this red noise for each lightcurve. This parameter can be readily recovered from plots like Fig.~\ref{magsig}. $\sigma_r$ is the value needed to attain the second sets of points from the third when quadratically added.
The results of this exercise were clear and very instructive: all the surveys examined are affected by significant correlated noise at the level of a few millimagnitudes. Typically, the red noise causes the standard deviation of the mean to be 2-3 times larger than expected for uncorrelated noise over transit-length timescales. 

One strong result is that sophisticated decorrelation algorithms like TFA and SysRem are very effective at reducing the red noise. Before their application, $\sigma_r$ usually amounts to 3-6 mmag for the best targets, while after it can be reduced to 1-2 mmag. 

We found no obvious relation between the depth of the survey and the amplitude of the red noise. Given that these surveys vary very widely in telescope size (from 10 cm to 4 m), in pixel scale and in field crowdedness, it is somewhat surprising to find little dependence in the final amount of red noise. 

There are, however,  two distinct behaviours for the red noise depending on the surveys: in some surveys the red noise is independent of magnitude, in others it is proportional to the photon noise. Among the six surveys that we examined, it seems that the ones using large telescopes show the first behaviour, and the ones using small cameras the second. This may be an indication that the correlated noise does not have the same underlying causes in deep and in shallow surveys.


The work of the ISSI group on this topic is still on-going, and our next objective is to identify the main causes of the residual correlated noise, and the source of the difference between its two observed behaviours.

One of the motivations in setting up the ISSI group was to determine which was the ``ideal'' instrumental setup for a photometric transit survey, i.e. in which regime the red noise could be minimised for the largest number of targets. The conclusion of our work is that there is to first order no preferred setup, with comparable results all along the magnitude scale. This may be due to the fact that ultimately all surveys operate under a similar atmosphere - the Earth's - and with similar detectors - large CCD cameras. Other parameters may be of smaller significance. 
As a consequence, it seems that surveys with targets of all magnitudes can contribute in comparable way to the detection of transiting planets.

There is a very important caveat to this, however. The scientific usefulness of a transiting planet is a very steep function of its magnitude. Transits on a bright star allow many kinds of subsequent studies that are not possible on fainter stars. For the faintest stars, as pointed out earlier, the planetary nature of the transit itself cannot be ascertained.

\section{``All right, then, how many will I find?''}

We have now come full circle and think we understand better the puzzling characteristics of the transiting planets found by photometric surveys, and the tenfold gap between some predictions and actual detection rates. Correlated noise in ground-based photometric data produces a marked increase in the detection threshold, so that only planets caught a large number of time in transit  and/or with especially deep transits can be detected. The number of measured transits is enhanced for shorter period, and for periods resonant with one day. Hence the features of the detected planets: very short periods, resonant periods, and transits deeper than average. Hence also the fact that many transiting hot Jupiters present among the targets were not detected.

\begin{figure}[!ht]
\centering
\includegraphics[angle=-90,width=7cm]{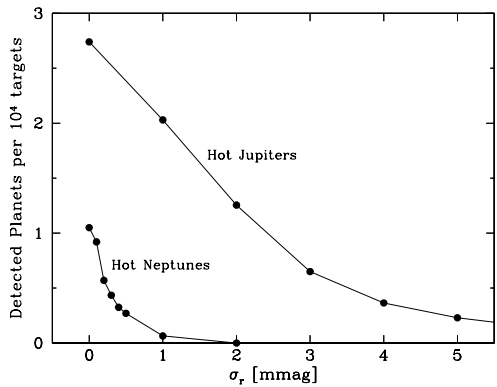}
\caption{Expected number of transiting planet detections for a representative transit survey, as a function of the level of red noise, for hot Jupiters (upper curve) and hot Neptunes.\label{HJHN}}
\end{figure}

Numerical simulations based on the PZQ modelling of the correlated noise allow more quantitative predictions. Fig.~\ref{HJHN} shows the dependence of the total catch in transiting hot Jupiters of a typical ground-based transit survey, as a function of the level of red noise characterised by $\sigma_r$. The potential planet catch is reduced by a factor 5-10 for typical ground-based photometric data before decorrelation. This goes a good way towards explaining the one-order-of-magnitude discrepancy mentioned earlier. The reduction factor is about 2 for the best lightcurves in ground-based surveys after decorrelation, a value that one may deem satisfactory, given the large effort probably involved in reducing it further. These calculations also show that with $\sigma_r$ above a few millimag, the efficiency of a transit search drops to negligible values.

Another implication is that an increased detection threshold will not only reduce the total planet catch, but also increase the selection effects, particularly in terms of period and planetary radius.

When  transits of smaller planets are considered, like hot Neptunes ($\sim $3 R$_\oplus$), the expected catch quickly leaves the white-noise regime and drops to zero. Extremely small amounts of correlated noise are necessary to detect hot Neptunes transiting solar-type stars from the ground by transit surveys (as proposed e.g. by Hartman et al. 2005 and Gillon et al. 2005).

Correlated noise is, of course, one of the main reasons for sending transit search missions like Corot and Kepler into space (the other reason being the access to continuous time coverage). Systematics will be reduced in space. However, they will not be irrelevant. On the contrary: calculations show that for the most interesting objects in the mission - the telluric planets - the photon noise will be sufficiently low for the red noise to determine the detection threshold. The control over transit-timescale correlation is an important factor controlling the minimum size of the planets that the space-born missions will be able to detect. Activity and granulation noise of the star itself may often prove to be the dominant red noise source. We hope that the efforts of the ISSI group to address the effect of correlated noise and to identify its causes will also be useful in that context.


\acknowledgements 
We thank the International Institute for Space Science in Bern, Switzerland, for providing the facilities and funding to set up the Transiting Planet Working Group.


\end{document}